\documentclass[
    ,final            % use final for the camera ready runs
  ]
  {aipproc}

\layoutstyle{8x11single}

\usepackage{graphicx}
\usepackage{latexsym}
\usepackage{amsfonts}
\usepackage{color}
\usepackage{colortbl}
\usepackage{amssymb,amsmath,enumerate,undertilde}
\usepackage{amsthm} % AMS package for THEOREM environment
\usepackage{bm}
\usepackage{indentfirst}
\usepackage{mathrsfs}
\usepackage[T1]{fontenc}
\usepackage{psfrag}

\newcommand{\Real}{\mathbb{R}}
\newcommand{\divrg}{\mbox{div}\,}

\newtheorem{remark}{Remark}
\newtheorem{lemma}{Lemma}
\newtheorem{proposition}{Proposition}

\newtheorem{theorem}{Theorem}

\begin{document}
%%%%%%%%%%%%%%%%%%%%%%%%%%%%%%%%%%%%%%%%%%%%

\title{Tangent bundle formulation of a charged gas}

\begin{abstract}
We discuss the relativistic kinetic theory for a simple, collisionless, charged gas propagating on an arbitrary curved spacetime geometry. Our general relativistic treatment is formulated on the tangent bundle of the spacetime manifold and takes advantage of its rich geometric structure. In particular, we point out the existence of a natural metric on the tangent bundle and illustrate its role for the development of the relativistic kinetic theory. This metric, combined with the electromagnetic field of the spacetime, yields an appropriate symplectic form on the tangent bundle. The Liouville vector field arises as the Hamiltonian vector field of a natural Hamiltonian. The latter also defines natural energy surfaces, called mass shells, which turn out to be smooth Lorentzian submanifolds.

A simple, collisionless, charged gas is described by a distribution function which is defined on the mass shell and satisfies the Liouville equation. Suitable fibre integrals of the distribution function define observable  fields on the spacetime manifold, such as the current density and stress-energy tensor. Finally, the geometric setting of this work allows us to discuss the relationship between the symmetries of the electromagnetic field, those of the spacetime metric, and the symmetries of the distribution function. Taking advantage of these symmetries, we construct the most general solution of the Liouville equation an a Kerr-Newman black hole background.
\end{abstract}

% 04.20.-q: Classical GR
% 04.20.Ex: Initial value problem, existence and uniqueness of solutions
% 04.25.-g: Approximation methods; equations of motion
% 04.25.D-: Numerical relativity
% 04.25.Nx: Post-Newtonian approximation; perturbation theory; related approximations
% 04.40.-b: Self-gravitating systems, continuous media and classical fields in curved spacetime
% 04.70.-s: Physics of black holes
% 05.20.Dd: Kinetic theory
% 97.60.Lf: Astronomy: Late stage of star evolution: black holes

\author{Olivier Sarbach and Thomas Zannias}{
address={Instituto de F\'{\i}sica y Matem\'aticas,
Universidad Michoacana de San Nicol\'as de Hidalgo\\
Edificio C-3, Ciudad Universitaria, 58040 Morelia, Michoac\'an, M\'exico.}
}

\maketitle

%%%%%%%%%%%%%%%%%%%%%%%%%%%%%%%%%%%%%%%%%%%%%
\section{Introduction}
%%%%%%%%%%%%%%%%%%%%%%%%%%%%%%%%%%%%%%%%%%%%%

Many astrophysical or cosmological configurations involve the interaction of charged particles and electromagnetic fields on a background gravitational field. This setting can describe configurations such as pulsar magnetospheres, accretion flows on black holes and the complex dynamics of  the primordial plasma. Moreover, the spectra of many active galactic nuclei or supernova remnants are interpreted as synchrotron radiation emitted by charged relativistic particles (electrons) gyrating around the magnetic field lines. Depending upon the prevailing physical conditions, the ideal MHD regime, formulated on a curved spacetime, provides a reliable description. In other scenarios, the overall electric neutrality of the fluid component is a poor approximation and thus methods of relativistic kinetic theory are becoming relevant. In particularly, the relativistic version of the Vlasov equation plays a central role for the description of the charged component.

In two recent articles~\cite{oStZ13,oStZ13b}, motivated by early work by Synge, Israel and Ehlers~\cite{jS34,Synge-Book,wI63,wI72,jE71,jE73}, we gave a mathematically oriented introduction to the relativistic kinetic theory of gases. In~\cite{oStZ13}, we developed the kinetic theory of a relativistic simple gas, that is a collection of neutral, spinless classical particles of the same positive rest mass $m > 0$. This development was based on ideas of  symplectic geometry and Hamiltonian dynamics and the starting point in this construction, was the Poincar\'e one-form $\Theta$ defined on the tangent bundle $TM$ associated with the spacetime $(M,g)$. This form induces a symplectic structure  on the tangent bundle and this combined with a natural Hamiltonian  gave rise to the Liouville vector field $L$  on $TM$. The Hamiltonian function defines suitable energy surfaces  $\Gamma_m$ on $TM$ referred to as mass shells, having the property that when $L$ is restricted to these mass shells, the projections of the integral curves of $L$ on the spacetime manifold $M$ define a family of future directed timelike geodesics. In this framework the gas is described by a distribution function $f$ defined on the associated mass shell $\Gamma_m$. For a simple, collisionless gas, the distribution function $f$ obeys the Liouville equation $L[f] = 0$ and leads to a set of  observables whose construction and properties are discussed in~\cite{oStZ13}. As a further application of the symplectic-Hamiltonian framework, the kinetic theory of a relativistic, simple, charged gas was also developed. For this system, a generalized Poincar\'e one-form that gets a contribution from the background electromagnetic field is sufficient for the development of the theory. 
 
In the second work~\cite{oStZ13b}, we presented an alternative formulation 
of relativistic kinetic theory which complements the symplectic-Hamiltonian approach in~\cite{oStZ13}. The starting point in~\cite{oStZ13b} was the splitting of the tangent space of $TM$ into horizontal and vertical subspaces induced by the Levi-Civita connection of the spacetime manifold $(M,g)$. This splitting leads to the presence of a natural metric $\hat{g}$ on $TM$ and an almost  complex structure which together lead to the introduction of the Liouville vector field $L$ which is horizontal by construction. The metric $\hat{g}$ and the Liouville vector field give rise to the Hamiltonian function $H:=\hat{g}(L,L)/2$ on $TM$ and associated mass shells $\Gamma_m$ which are Lorentzian submanifolds of $TM$. Accordingly, there is a natural volume form on $\Gamma_m$ which allows us to define integrals of functions on the mass shell. In this approach the distribution function $f$ associated to a simple gas is again defined on the mass shell $\Gamma_m$, but presently in combination with the Liouville vector field $L$ can be viewed as describing a fictitious incompressible fluid on $\Gamma_m$ with associated current density ${\cal J} = f L/m$. For a collisionless gas ${\cal J}$ is divergence-free and it follows from Liouville's theorem that the distribution functions $f$ obeys the Liouville equation $L[f] = 0$. Furthermore, the current density ${\cal J}$ on $\Gamma_m$ gives rise to a physical current density $J$  on the spacetime manifold $(M,g)$. This physical current represents the first moment of the distribution function through a fibre integral. Higher moments of the distribution function can be constructed in an analogous way and shown to be conserved as well. Of particular relevance is the second moment which gives rise to the stress-energy tensor that allows one to couple gravity to the kinetic gas through Einstein's field equations.
  
The insights gained from this geometric formulation of kinetic theory are helpful in various aspects. Primarily, they lead to a clear understanding of the relationship between the symmetries of the spacetime manifold $(M,g)$ and those of the distribution function $f$. This connection has been exploited in~\cite{oStZ13b} where the most general spherically symmetric distribution function on an arbitrary spherically symmetric spacetime manifold has been discussed. Moreover, in~\cite{oStZ13b}, the most general solution of the Liouville equation on a Kerr black hole background has also been derived.

In view of the benefits and the complementary insights that the geometrical approach in~\cite{oStZ13b} offers in the description of the kinetic theory, in this paper we employ this approach to discuss the kinetic theory of  relativistic simple charged gases. Although for such cases, the spacetime $(M,g)$ contains a nonvanishing electromagnetic field $F$ and thus the gas particles do not any longer move on the geodesics of the background spacetime, nevertheless the kinetic theory can be deduced again from the splitting of the tangent space of $TM$ into horizontal and vertical subspaces induced by the Levi-Civita connection of the spacetime manifold $(M,g)$. In particular, the bundle metric $\hat{g}$ is not affected by the electromagnetic field $F$. However, the symplectic form is modified by the pull-back $\pi^* F$ of $F$ with respect to the natural projection $\pi: TM\to M$. This symplectic form, combined with the same Hamiltonian as in the uncharged case, gives rise to the Liouville vector field $L_F$ which now contains a vertical component related to the electromagnetic field $F$. As for the uncharged case, the Hamiltonian $H$ defines the mass shell $\Gamma_m$  on which the integral curves of $L_F$ are restricted. The  bundle metric $\hat{g}$ induces a Lorentzian metric on $\Gamma_m$ on which the Liouville vector field $L_F$ is divergence-free. The distribution function $f$ is defined as a nonnegative function on $\Gamma_m$, which together with $L_F$, can be thought of as defining a fictitious incompressible fluid on $\Gamma_m$ with current density ${\cal J}_F = f L_F/m$. As we discuss, the splitting of ${\cal J}_F$ into horizontal and vertical components resembles the familiar decomposition of the current density into advection and conduction currents. For a simple, collisionless, charged gas the distribution function satisfies the Liouville (or Vlasov) equation $L_F[f] = 0$. As for the uncharged case, suitable fibre integrals of the distribution function define observable  fields on the spacetime manifold such as the current density and stress-energy tensor. Finally, the geometric setting of this work allows us to discuss the relationship between symmetries of the electromagnetic field, those of the spacetime metric $g$ and symmetries of the distribution function $f$. As an application of this analysis, we discuss the Liouville equation $L_F[f] = 0$ on a Kerr-Newman background describing a charged, rotating black hole configuration. Based on symmetry considerations and the separability of the Hamilton-Jacobi equation, we show that the Liouville vector field can be trivialized by means of a suitable symplectic transformation on the tangent bundle and this result generalize the method discussed in Ref.~\cite{oStZ13b} valid for the case of uncharged gases.  

As in the previous papers~\cite{oStZ13,oStZ13b}, we develop the theory assuming a spacetime $(M,g)$ of an arbitrary dimension $n\geq 2$ and use the same notations and conventions as in~\cite{oStZ13,oStZ13b}. Specifically, $(M,g)$ denotes a $C^\infty$-differentiable, $n$-dimensional Lorentzian manifold with the signature convention $(-,+,+,\ldots,+)$ for the metric. We use the Einstein summation convention with Greek indices $\mu,\nu,\sigma,\ldots$ running from $0$ to $d=n-1$ and Latin indices $i,j,k,\ldots$ running from $1$ to $d$. For any $C^\infty$- differentiable manifold $N$, ${\cal X}(N)$ denotes the class of $C^\infty$-differentiable vector fields. The operators $i_X$ and $\pounds_X$ refer to the interior product and Lie derivative,  respectively, with respect to the vector field $X$. Round brackets enclosing indices refer to total symmetrization, for example $v_{(ij)} := (v_{ij} + v_{ji})/2$. We use units for which $c=1$.

%%%%%%%%%%%%%%%%%%%%%%%%%%%%%%%%%%%%%%%%%%%%%
\section{Geometry of the tangent bundle}
%%%%%%%%%%%%%%%%%%%%%%%%%%%%%%%%%%%%%%%%%%%%%

In this section, we briefly review the basic geometric properties of the tangent bundle $TM$ of an arbitrary, $n$-dimensional spacetime manifold $(M,g)$. In particular, we discuss the splitting of the tangent space at any point of $TM$ into a horizontal and a vertical subspace and the canonical bundle metric $\hat{g}$ on $TM$ induced by this splitting. Moreover, we introduce an appropriate symplectic form on $TM$. As it turns out, both of these structures provide the building blocks for the formulation of relativistic kinetic theory. Specifically, the metric allows us to define an integration theory on submanifolds of $TM$ which is essential for the invariant definition of the distribution function. The symplectic form provides the means for the Hamiltonian formulation of the theory.

Let $T_x M$ denote the vector space of all tangent vectors $p$ at some event $x\in M$. The tangent bundle of $M$ is defined as
\begin{displaymath}
TM := \{ (x,p) : x\in M, p\in T_x M \},
\end{displaymath}
with the associated projection map $\pi: TM\to M$, $(x,p)\mapsto x$. The fibre at $x\in M$ is the space $\pi^{-1}(x) = (x,T_x M)$ which is naturally  isomorphic to $T_x M$. The first basic property of the tangent bundle is described in the following lemma.

\begin{lemma}
\label{Lem:TM}
$TM$ is an orientable, $2n$-dimensional $C^\infty$-differentiable manifold.
\end{lemma}

For a proof, see Ref.~\cite{DoCarmo-Book1} or Refs.~\cite{oStZ13,oStZ13b}. The idea is to start with local coordinates $(x^\mu)$ on an open subset $U$ of $M$. Then, we associate to each point $(x,p)\in V := \pi^{-1}(U)$ the coordinates $(x^\mu,p^\mu)$ with $x^\mu$ the coordinates of $x$ and $p^\mu := dx_x^\mu(p)$. We call $(x^\mu,p^\mu)$ \emph{adapted local coordinates}; the associated basis of the tangent and cotangent spaces of $TM$ at $(x,p)\in V$ are
\begin{displaymath}
\left\{ \left. \frac{\partial}{\partial x^\mu} \right|_{(x,p)},\left. \frac{\partial}{\partial p^\mu} \right|_{(x,p)} \right\},\qquad
\left\{ dx^\mu_{(x,p)},dp^\mu_{(x,p)} \right\}.
\end{displaymath}

\subsection{Splitting into horizontal and vertical subspaces}

For a given spacetime manifold $(M,g)$ there exist two natural projection maps $T_{(x,p)}(TM)\to T_x M$ which assign to each tangent vector $X\in T_{(x,p)}(TM)$ a unique tangent vector in $T_x M$. These maps define a natural splitting of the tangent space $T_{(x,p)}(TM) = H_{(x,p)}\oplus V_{(x,p)}$ at any point $(x,p)\in TM$, where $H_{(x,p)}$ and $V_{(x,p)}$ are called the horizontal and vertical subspace, respectively.

The first projection map arises through the push-forward of the map $\pi: TM\to M$, which induces the map $\pi_{*(x,p)}: T_{(x,p)}(TM)\to T_x M$. It is a simple matter to verify that in adapted local coordinates $(x^\mu,p^\mu)$ we have
\begin{displaymath}
\pi_{*(x,p)}\left( X^\mu(x,p)\left. \frac{\partial}{\partial x^\mu} \right|_{(x,p)}
+ Y^\mu(x,p)\left. \frac{\partial}{\partial p^\mu} \right|_{(x,p)} \right) 
 = X^\mu(x,p)\left. \frac{\partial}{\partial x^\mu} \right|_x
\end{displaymath}
for any vector field $Z = X^\mu\frac{\partial}{\partial x^\mu} + Y^\mu\frac{\partial}{\partial p^\mu}$ on $TM$. At any point $(x,p)\in TM$, the \emph{vertical subspace} $V_{(x,p)}$ of $T_{(x,p)}(TM)$ is defined as the following $n$-dimensional subspace of $T_{(x,p)}(TM)$:
\begin{equation}
V_{(x,p)} := \ker \pi_{*(x,p)} = \{ Z\in T_{(x,p)}(TM) : \pi_{*(x,p)}(Z) = 0 \}.
\end{equation}
In terms of adapted local coordinates $V_{(x,p)}$ is generated by the vectors $\left. \frac{\partial}{\partial p^\mu} \right|_{(x,p)}$, $\mu = 0,1,\ldots d$.

\begin{figure}[ht]
\centerline{\resizebox{9.0cm}{!}{\includegraphics{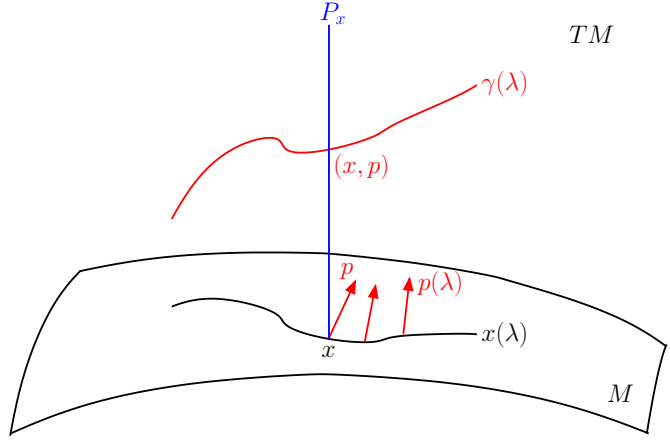}}}
\caption{The curve $\gamma(\lambda)$ on $TM$ which gives rise to the curve $x(\lambda)$ on $M$ and the vector field $p(\lambda)$ along it.}
\label{Fig:Connection}
\end{figure}

The second projection map $K_{(x,p)}: T_{(x,p)}(TM)\to T_x M$, called the \emph{connection map} (see Refs.~\cite{pD62,sGeK02}), makes use of the Levi-Civita connection $\nabla$ of the background spacetime $(M,g)$ and is defined as follows: let $\gamma(\lambda) = (x(\lambda),p(\lambda))$ be a smooth curve in $TM$ through $(x,p)$ with tangent vector $Z$ at $(x,p)$, that is, $\gamma(0) = (x,p)$ and $\dot{\gamma}(0) = Z$. The curve $\gamma(\lambda)$ gives rise to the curve $x(\lambda)$ in $M$ and the vector field $p(\lambda)$ along it, see Figure~\ref{Fig:Connection}. Then, we define
\begin{equation}
K_{(x,p)}(Z) 
 := \left. \frac{d}{d\lambda} \tau_{0,\lambda} p(\lambda) \right|_{\lambda=0}
 = (\nabla_{\dot{x}(0)} p)_x,
\label{Eq:KDef}
\end{equation}
where $\tau_{0,\lambda}: T_{x(\lambda)} M\to T_x M$ denotes the parallel transport operator along $x(\lambda)$. In terms of adapted local coordinates $(x^\mu,p^\mu)$ one can show that
\begin{displaymath}
K_{(x,p)}(Z) = \left[ Y^\mu 
 + \Gamma^\mu{}_{\alpha\beta}(x) X^\alpha p^\beta \right]
 \left. \frac{\partial}{\partial x^\mu} \right|_x,\qquad
 Z = X^\mu\left. \frac{\partial}{\partial x^\mu} \right|_{(x,p)}
 + Y^\mu\left. \frac{\partial}{\partial p^\mu} \right|_{(x,p)},
\end{displaymath}
where $\Gamma^\mu{}_{\alpha\beta}$ denote the Christoffel symbols.

It is a simple matter to verify the following Lemma.

\begin{lemma}
\label{Lem:ConnectionMap}
The connection map $K_{(x,p)} : T_{(x,p)}(TM)\to T_x M$ satisfies:
\begin{enumerate}
\item[(i)] $K_{(x,p)}$ is a linear map.
\item[(ii)] $H_{(x,p)} := \ker K_{(x,p)}$ is a $n$-dimensional subspace of $T_{(x,p)}(TM)$.
\item[(iii)] $H_{(x,p)} \cap V_{(x,p)} = \{ 0 \}$.
\end{enumerate}
\end{lemma}

The horizontal space $H_{(x,p)}$ is generated by the following $n$ tangent vectors:
\begin{equation}
\left. e_\mu \right|_{(x,p)} := \left. \frac{\partial}{\partial x^\mu} \right|_{(x,p)} 
 - \Gamma^\alpha{}_{\mu\beta}(x) p^\beta
 \left. \frac{\partial}{\partial p^\alpha} \right|_{(x,p)},\qquad
\mu = 0,1,\ldots d,
\label{Eq:emu}
\end{equation}
so that $\{ \left. e_\mu \right|_{(x,p)}, \left. \frac{\partial}{\partial p^\mu} \right|_{(x,p)} \}$ is a basis of $T_{(x,p)}(TM)$ adapted to the splitting $T_{(x,p)}(TM) = H_{(x,p)}\oplus V_{(x,p)}$. The corresponding dual basis is given by $\{ dx^\mu_{(x,p)}, \theta^\mu_{(x,p)} \}$ with
\begin{equation}
\theta^\mu_{(x,p)} 
 := dp^\mu_{(x,p)} + \Gamma^\mu{}_{\alpha\beta}(x) p^\beta dx^\alpha_{(x,p)},
 \qquad \mu = 0,1,\ldots d.
\label{Eq:thetamu}
\end{equation}
The vector fields $e_\mu$ and $\frac{\partial}{\partial p^\mu}$ satisfy the following commutation relations which will be useful later:
\begin{equation}
[e_\mu,e_\nu] = -R^\alpha{}_{\beta\mu\nu} p^\beta\frac{\partial}{\partial p^\alpha},
\quad
\left[ e_\mu, \frac{\partial}{\partial p^\nu} \right] = \Gamma^\alpha{}_{\mu\nu}
\frac{\partial}{\partial p^\alpha},
\quad
\left[ \frac{\partial}{\partial p^\mu} , \frac{\partial}{\partial p^\nu} \right] = 0,
\label{Eq:ComRel}
\end{equation}
where $R^\alpha{}_{\beta\mu\nu}$ denotes the curvature tensor on $(M,g)$.

As a consequence of Lemma~\ref{Lem:ConnectionMap}, any tangent vector $Z\in T_{(x,p)}(TM)$ can be uniquely decomposed as
\begin{equation}
Z = Z^H + Z^V,\qquad Z^H\in H_{(x,p)},\quad Z^V\in V_{(x,p)},
\label{Eq:HorVerSplit}
\end{equation}
where the horizontal and vertical components can be written as
\begin{equation}
Z^H = X^\mu\left. e_\mu \right|_{(x,p)},\quad  
Z^V =  Y^\mu\left. \frac{\partial}{\partial p^\mu} \right|_{(x,p)},
\end{equation}
with $X^\mu = dx^\mu_{(x,p)}(Z)$ and $Y^\mu = \theta^\mu_{(x,p)}(Z)$. The natural splitting of the tangent space into horizontal and vertical subspaces yield the following linear isomorphisms:
\begin{eqnarray}
&& I^H_{(x,p)} := \left. \pi_{*(x,p)} \right|_{H_{(x,p)}} : H_{(x,p)} \to T_x M,
\label{Eq:IH}\\
&& I^V_{(x,p)} := \left. K_{(x,p)} \right|_{V_{(x,p)}} : V_{(x,p)} \to T_x M,
\label{Eq:IV}
\end{eqnarray}
which allows us to identify each of the two spaces $H_{(x,p)}$ and $V_{(x,p)}$ with the tangent space $T_x M$. As a consequence, we can introduce an almost complex structure on $TM$ which rotates horizontal vectors into vertical ones and vice versa. More specifically, it is defined as the linear map $J_{(x,p)} : T_{(x,p)}(TM) \to T_{(x,p)}(TM)$, satisfying $J^2 = -1$, given by
\begin{equation}
J_{(x,p)}(Z) = J_{(x,p)}(Z^H + Z^V)
 := (I^V_{(x,p)})^{-1}\circ I^H_{(x,p)}(Z^H) 
 - (I^H_{(x,p)})^{-1}\circ I^V_{(x,p)}(Z^V)
\label{Eq:AlmostComplexStruc}
\end{equation}
for all $Z\in T_{(x,p)}(TM)$. In terms of the basis vectors $\{ \left. e_\mu \right|_{(x,p)} , \left. \frac{\partial}{\partial p^\mu} \right|_{(x,p)} \}$ of $T_{(x,p)}(TM)$ the maps $I^H_{(x,p)}$, $I^V_{(x,p)}$ and $J_{(x,p)}$ have the following representations:
\begin{eqnarray}
&& I^H_{(x,p)}\left( X^\mu\left. e_\mu \right|_{(x,p)} \right) = 
X^\mu\left. \frac{\partial}{\partial x^\mu} \right|_x ,
\label{Eq:IHLocCoord}\\
&& I^V_{(x,p)}\left( Y^\mu\left. \frac{\partial}{\partial p^\mu} \right|_{(x,p)} \right) = 
Y^\mu\left. \frac{\partial}{\partial x^\mu} \right|_x ,
\label{Eq:IVLocCoord}\\
&& J_{(x,p)}\left( X^\mu\left. e_\mu \right|_{(x,p)} 
 + Y^\mu\left. \frac{\partial}{\partial p^\mu} \right|_{(x,p)} \right)
 =  -Y^\mu\left. e_\mu \right|_{(x,p)} 
 + X^\mu\left. \frac{\partial}{\partial p^\mu} \right|_{(x,p)}.
\label{Eq:JLocCoord}
\end{eqnarray}
As seen from these expressions, the operators $I^H_{(x,p)}$, $I^V_{(x,p)}$, $J_{(x,p)}$ are smooth in $(x,p)$.

\subsection{The canonical bundle metric}

Based on the splitting of the tangent space of $TM$ into horizontal and vertical subspaces, the spacetime metric $g$ induces a natural metric $\hat{g}$ on $TM$. In the context of Riemannian manifolds the metric $\hat{g}$ was introduced a long time ago by Sasaki~\cite{sS58}. Its relevance for the description of relativistic kinetic theory was pointed out in~\cite{oStZ13b}, where a simple uncharged gas was treated. In this work we show that the metric $\hat{g}$ also plays an important role in the description of a simple charged gas.

In terms of the Lorentzian metric $g$ on the base manifold, the push-forward of $\pi$ and the connection map $K$ the metric $\hat{g}$ is defined as
\begin{equation}
\hat{g}_{(x,p)}(Z,W)  := g_x( \pi_{*(x,p)}(Z), \pi_{*(x,p)}(W) ) + g_x( K_{(x,p)}(Z), K_{(x,p)}(W) ),
\label{Eq:BundleMetric}
\end{equation}
for all $Z,W\in T_{(x,p)}(TM)$. In terms of the basis covectors $\{ \left. dx^\mu \right|_{(x,p)} , \left. \theta^\mu \right|_{(x,p)} \}$ of $T_{(x,p)}^*(TM)$ defined in Eq.~(\ref{Eq:thetamu}) the canonical bundle metric $\hat{g}$ has the representation
\begin{equation}
\hat{g}_{(x,p)} = g_{\mu\nu}(x) dx^\mu_{(x,p)}\otimes dx^\nu_{(x,p)}
 + g_{\mu\nu}(x)\theta^\mu_{(x,p)}\otimes\theta^\nu_{(x,p)},
\label{Eq:BundleMetricCoord}
\end{equation}
where $g_{\mu\nu}(x)$ are the coordinate components of $g$. The most important properties of this metric are summarized in the following Lemma.

\begin{lemma}
The canonical bundle metric $\hat{g}$ is a semi-Riemannian metric satisfying:
\begin{enumerate}
\item[(i)] The signature of $\hat{g}$ is $(2,2n-2)$.
\item[(ii)] The splitting in Eq.~(\ref{Eq:HorVerSplit}) is orthogonal with respect to $\hat{g}$, that is, $\hat{g}_{(x,p)}(Z^H,Z^V) = 0$ for all $Z^H\in H_{(x,p)}$ and $Z^V\in V_{(x,p)}$.
\item[(iii)] The metric $\hat{g}$ is invariant with respect to the almost complex structure $J$ defined in Eq.~(\ref{Eq:AlmostComplexStruc}): $\hat{g}_{(x,p)}(J(Z),J(W)) = \hat{g}_{(x,p)}(Z,W)$ for all $Z,W\in T_{(x,p)}(TM)$.
\end{enumerate}
\end{lemma}

For later use, we note that the bundle metric $\hat{g}$ induces a natural volume form on the tangent bundle $TM$, given by
\begin{equation}
\eta_{TM} := -\det(g_{\mu\nu})
dx^0\wedge \ldots\wedge dx^d\wedge
dp^0\wedge \ldots\wedge dp^d.
\label{Eq:VolumeFormTM}
\end{equation}
For further properties of the Sasaki metric we refer the reader to~\cite{sS58,sGeK02,oStZ13b}.

\subsection{The symplectic form}

So far, all the results regarding the geometry of the tangent bundle were the natural outcomes of the invariant splitting of the tangent space of the tangent bundle, which in turn relied on the metric and the associated Levi-Civita connection on the base manifold $M$. For the next result, we shall make use also of a closed two-form field $F$ on the base manifold $M$. This field represents a background electromagnetic field on $M$ and will play an important role in the description of charged relativistic gases. Specifically, $g$ and $F$ define a particular symplectric two-form $\Omega_F$ on $TM$. Recalling the almost complex structure $J$ defined in Eq.~(\ref{Eq:AlmostComplexStruc}), we define $\Omega_F$ as
\begin{equation}
\Omega_F(Z,W) := \hat{g}(Z, J(W)) + q(\pi^* F)(Z,W),
\label{Eq:OmegaF}
\end{equation}
for two vector fields $Z$ and $W$ on $TM$, where $\pi^* F$ denotes the pull-back of $F$ with respect to the projection map $\pi: TM\to M$ and $q$ is a constant representing the charge of the gas particles. In terms of the basis covectors $\{ \left. dx^\mu \right|_{(x,p)} , \left. \theta^\mu \right|_{(x,p)} \}$ of $T_{(x,p)}^*(TM)$ defined in Eq.~(\ref{Eq:thetamu}) the symplectic form can be written as
\begin{equation}
\Omega_F = g_{\mu\nu}\theta^\mu\wedge dx^\nu 
 + \frac{q}{2} F_{\mu\nu} dx^\mu\wedge dx^\nu
 = g_{\mu\nu} dp^\mu\wedge dx^\nu 
 + \frac{\partial g_{\mu\nu}}{\partial x^\alpha} p^\mu dx^\alpha\wedge dx^\nu
 + \frac{q}{2} F_{\mu\nu} dx^\mu\wedge dx^\nu.
\label{Eq:OmegaFLocCoord}
\end{equation}
Using the properties of the bundle metric $\hat{g}$ it is not difficult to check that $\Omega_F$ is antisymmetric and non-degenerated. Moreover, $\Omega_F$ is closed since $\Omega_F = d\Theta_A$ where the one-form $\Theta_A$ on the tangent bundle is defined as~\cite{oStZ13}
\begin{equation}
\Theta_{A(x,p)}(X) := g_x(p,\pi_{*(x,p)}(X)) + q A_x(\pi_{*(x,p)}(X))
\label{Eq:PoincareOneForm}
\end{equation}
for $X\in T_{(x,p)}(TM)$, with $A$ a vector potential such that $F = dA$. It should be mentioned that in the uncharged case $q=0$ the symplectic form in Eq.~(\ref{Eq:OmegaF}) reduces to the form $\Omega_s$ introduced in Ref.~\cite{oStZ13,oStZ13b}, and the one-form $\Theta_A$ reduces to the Poincar\'e one-form on $TM$.

In view of Liouville's theorem below, an important property of the symplectic form $\Omega_F$ is provided by the following relation:

\begin{lemma}
\label{Lem:VolumeTM}
Let $\eta_{TM}$ be the volume form on $TM$ induced by the bundle metric $\hat{g}$, see Eq.~(\ref{Eq:VolumeFormTM}). Then,
\begin{equation}
\eta_{TM} = -\frac{(-1)^{\frac{n(n+1)}{2}}}{n!} 
\Omega_F\wedge\Omega_F\wedge\ldots\wedge\Omega_F,\qquad
\hbox{($n$-fold product)}.
\end{equation}
\end{lemma}

\proof As in the uncharged case, see Ref.~\cite{oStZ13b}.
\qed

%%%%%%%%%%%%%%%%%%%%%%%%%%%%%%%%%%%%%%%%%%%%%
\section{The kinetic theory for a simple, collisionless, charged gas}
\label{Sec:KTheory}
%%%%%%%%%%%%%%%%%%%%%%%%%%%%%%%%%%%%%%%%%%%%%

After discussing the geometrical aspects of the tangent bundle, in this section we apply this framework to the description of relativistic kinetic theory for a simple charged gas, that is, a collection of classical, spinless particles of the same positive mass $m$ and the same charge $q$. Following the same construction as in Ref.~\cite{oStZ13b} for the uncharged case, we first introduce the Liouville vector field $L_F$ and a suitable Hamiltonian function $H$ on the tangent bundle $TM$. The integral curves of $L_F$ are related to the possible trajectories of the gas particles between collisions. The Hamiltonian $H$ generates these integral curves via the symplectic form $\Omega_F$ defined in the previous section. Moreover, the Hamiltonian defines the mass shell $\Gamma_m$ on which the integral curves of $L_F$ are restricted. The bundle metric $\hat{g}$ induces a Lorentzian metric on $\Gamma_m$ on which the Liouville vector field $L_F$ is divergence-free. Consequently, $L_F$ generates an incompressible flow on $\Gamma_m$ which in turn implies Liouville's theorem. The distribution function is defined as a nonnegative function $f: \Gamma_m \to \Real$, which together with $L_F$, can be thought of as defining a fictitious incompressible fluid on $\Gamma_m$ with current density ${\cal J}_F = f L_F/m$. For a collisionless gas ${\cal J}_F$ is divergence-free and Liouville's equation $L_F[f] = 0$ follows.

In contrast to the uncharged case, the Liouville vector field $L_F$ adquires a nontrivial vertical component in addition to the horizontal one, as we show below.

\subsection{Hamiltonian function and Liouville vector field}

The Hamiltonian function for our system is defined as
\begin{equation}
H(x,p) := \frac{1}{2} g_x(p,p),\qquad (x,p)\in TM.
\label{Eq:HDef}
\end{equation}
In conjunction with the symplectic form $\Omega_F$ this $H$ defines the associated Hamiltonian vector field $L_F$ through
\begin{equation}
dH = -i_{L_F}\Omega_F = \Omega_F(\cdot,L_F).
\label{Eq:LHam}
\end{equation}

\begin{lemma}
\label{Lem:LF}
The vector field $L_F$ is given by
\begin{equation}
L_{F(x,p)} = (I^H_{(x,p)})^{-1}(p) + q(I^V_{(x,p)})^{-1}(\tilde{F}_x(p)),
\label{Eq:LF}
\end{equation}
where $\tilde{F}: {\cal X}(M)\to {\cal X}(M)$ is defined by $g(X,\tilde{F}(Y)) = F(X,Y)$ for all $X,Y\in {\cal X}(M)$.
\end{lemma}

\proof This can be shown by a straightforward generalization of Lemma 6 in Ref.~\cite{oStZ13b}. Alternatively, using adapted local coordinates $(x^\mu,p^\mu)$, it follows from Eq.~(\ref{Eq:HDef}) that
\begin{equation}
dH = \frac{1}{2} d\left( g_{\mu\nu} p^\mu p^\nu \right)
 = g_{\mu\nu} p^\mu dp^\nu
 + \frac{1}{2}\frac{\partial g_{\mu\nu}}{\partial x^\alpha} p^\mu p^\nu dx^\alpha
 = g_{\mu\nu} p^\mu \theta^\nu,
\label{Eq:dH}
\end{equation}
where $\theta^\nu$ is defined in Eq.~(\ref{Eq:thetamu}). On the other hand, from Eq.~(\ref{Eq:LF}),
\begin{equation}
L_F = p^\mu e_\mu + q F^\mu{}_\nu p^\nu \frac{\partial}{\partial p^\mu}
 = p^\mu\frac{\partial}{\partial x^\mu}
 + \left[ q F^\mu{}_\nu p^\nu - \Gamma^\mu{}_{\alpha\beta} p^\alpha p^\beta \right]
 \frac{\partial}{\partial p^\mu},
\label{Eq:LLocCoord}
\end{equation}
where we have used the definition of $e_\mu$ in Eq.~(\ref{Eq:emu}). Using the coordinate expression~(\ref{Eq:OmegaFLocCoord}) for the symplectic form $\Omega_F$ we find from this
\begin{equation}
\Omega_F(\cdot,L_F) 
 = g_{\mu\nu}\left[ \theta^\mu dx^\nu(L_F) - dx^\nu \theta^\mu(L_F)\right]
 + q F_{\mu\nu} dx^\mu dx^\nu(L_F) 
 = g_{\mu\nu}\theta^\mu p^\nu = dH,
\end{equation}
which proves the lemma.
\qed

From the explicit representation in Eq.~(\ref{Eq:LF}) it follows that any integral curve $(x(\lambda),p(\lambda))$ of $L_F$ through the point $(x,p)$ satisfies
\begin{equation}
\nabla_p p = q\tilde{F}(p),\qquad
p = \dot{x}(0).
\end{equation}
Therefore, the projected curve $x(\lambda)$ on $M$ defines a dynamical trajectory of a charged particle on the spacetime manifold $(M,g)$ in the presence of the electromagnetic field $F$.

\subsection{The mass shell and its basic properties}

The Hamiltonian function gives rise to the mass shells, which are defined by
\begin{equation}
\Gamma_m := H^{-1}\left( -\frac{1}{2}m^{2} \right)
 = \{ (x,p)\in TM : g_x(p,p) = -m^2 \}.
\end{equation}
This set defines the energy surface in which the motion of particles of mass $m$ takes place. Since $H$ does not depend on the electromagnetic field $F$, it follows that $\Gamma_m$ satisfies the same properties as in the uncharged case:

\begin{proposition}
\label{Prop:MassShell}
Let $m\neq 0$. The set $\Gamma_m$ satisfies the following properties:
\begin{enumerate}
\item[(i)] $\Gamma_m$ is a $(2n-1)$-dimensional $C^\infty$-differentiable manifold.
\item[(ii)] Let $\hat{h}$ be the metric on $\Gamma_m$ induced by the bundle metric $\hat{g}$. Then $(\Gamma_m,\hat{h})$ is a Lorentz manifold with unit vector field $N = (I^V)^{-1}(p)/m$.
\item[(iii)] Suppose $M$ is connected, then $(M,g)$ is time-orientable if and only if $\Gamma_m$ is disconnected, in which case it is the disjoint union of two connected components $\Gamma_m^+$ and $\Gamma_m^-$.
\end{enumerate}
\end{proposition}

\proof See Lemma~5 and~7 in Ref.~\cite{oStZ13b} and Appendix A in Ref.~\cite{oStZ13}.
\qed

\begin{remark}
For the massless case $m=0$ the set $\Gamma_m$ is smooth except at the vertex points $p=0$.
\end{remark}

\begin{remark}
It follows from property (ii) that any horizontal vector field is tangent to $\Gamma_m$.
\end{remark}

\begin{remark}
For the following, we shall assume that $(M,g)$ is connected and time-oriented. In this case the ``future'' mass shell can also be written as $\Gamma_m^+ = \{ (x,p) : x\in M, p\in P_x^+ \}$, with the future mass hyperboloids
\begin{equation}
P_x^+ := \{ p\in T_x M : g_x(p,p) = -m^2, \hbox{$p$ future directed} \}.
\label{Eq:DefPx}
\end{equation}
Physically, the restriction on $\Gamma_m^+$ incorporates the idea that the gas particles move on future directed timelike curves.
\end{remark}

A local representation of $\Gamma_m^+$ and $\hat{h}$ can be obtained in the following way: Let $(U,\phi)$ be a local chart of $(M,g)$ with corresponding local coordinates $(x^0,x^1,\ldots,x^d)$, such that for each $x\in U$, $\left. \frac{\partial}{\partial x^0} \right|_x$ is timelike and all the vectors of the form $\left. p^i\frac{\partial}{\partial x^i} \right|_x$ are spacelike. Let $(V,\psi)$ denote the local chart of $TM$ with the corresponding adapted local coordinates $(x^\mu,p^\mu)$, see the comments below Lemma~\ref{Lem:TM}. Relative to these local coordinates, the mass shell is determined by
\begin{equation}
-m^2 = g_{\mu\nu}(x) p^\mu p^\nu 
 = g_{00}(x)(p^0)^2 + 2 g_{0j}(x) p^0 p^j + g_{ij}(x) p^i p^j.
\label{Eq:MassShellCond}
\end{equation}
Therefore, the future mass shell $\Gamma_m^+$ can be locally represented as those $(x^\mu,p^0,p^i)\in \psi(V)\subset \Real^{2n}$ for which $p^0 = p_+^0(x^\mu,p^i)$ with
\begin{equation}
p_+^0(x^\mu,p^i) := \frac{g_{0j}(x) p^j +
 \sqrt{[g_{0j}(x) p^j]^2 + [-g_{00}(x)]\left[ m^2 + g_{ij}(x) p^i p^j \right] }}
 {-g_{00}(x)} > 0.
\label{Eq:p0up}
\end{equation}
In terms of the resulting local coordinates $(x^\mu,p^i)$ on $\Gamma_m^+$, a basis of tangent and co-tangent vectors adapted to the splitting~(\ref{Eq:HorVerSplit}) is given by
\begin{equation}
\hat{e}_\mu := \frac{\partial}{\partial x^\mu} 
 - \Gamma^k{}_{\mu\beta}\hat{p}^\beta\frac{\partial}{\partial p^k},\qquad
\frac{\partial}{\partial p^i},
\label{Eq:ehatmu}
\end{equation}
and
\begin{equation}
dx^\mu,\qquad
\hat{\theta}^i := dp^i + \Gamma^i{}_{\alpha\beta}\hat{p}^\beta dx^\alpha
\end{equation}
for $\mu = 0,1,\ldots d$ and $i=1,\ldots d$, where we have defined $\hat{p}^0 := p_+^0(x^\mu,p^j)$ and $\hat{p}^j := p^j$. Using this notation, the induced metric $\hat{h}$ in the local coordinates $(x^\mu, p^i)$ has the form~\cite{oStZ13b}
\begin{equation}
\hat{h} = g_{\mu\nu} dx^\mu\otimes dx^\nu 
 + \left( g_{ij} - \frac{2}{\hat{p}_0} g_{0(i}\hat{p}_{j)} 
  + \frac{1}{\hat{p}_0^2} g_{00}\hat{p}_i\hat{p}_j\right)
\hat{\theta}^i\otimes\hat{\theta}^j,
\end{equation}
where here $\hat{p}_i := g_{i0}(x)\hat{p}^0 + g_{ij}(x)\hat{p}^j$, and 
\begin{equation}
\hat{p}_0 := p_{+ 0}(x^\mu,p^i) 
 = -\sqrt{ [g_{0j}(x) p^j]^2 + [-g_{00}(x)]\left[ m^2 + g_{ij}(x) p^i p^j \right]}.
\label{Eq:p0down}
\end{equation}
The volume form $\eta_{\Gamma_m}$ induced by the metric $\hat{h}$ can be written as~\cite{oStZ13b}
\begin{equation}
\eta_{\Gamma_m} = \frac{m}{\hat{p}_0}\det(g_{\mu\nu})
dx^0\wedge dx^1\wedge\ldots\wedge dx^d\wedge
dp^1\wedge \ldots\wedge dp^d.
\label{Eq:VolumeFormMassShellLocalCoord}
\end{equation}
Finally, we note that the Liouville vector field is tangent to the mass shell $\Gamma_m$ since $dH(L_F) = \Omega_F(L_F,L_F) = 0$. Therefore, we may also regard $L_F$ as a vector field on $\Gamma_m$. In terms of the local coordinates $(x^\mu,p^i)$ it has the form
\begin{equation}
L_F = \hat{p}^\mu\hat{e}_\mu + q F^i{}_\nu\hat{p}^\nu\frac{\partial}{\partial p^i}
 = \hat{p}^\mu\frac{\partial}{\partial x^\mu}
 + \left[ q F^i{}_\nu\hat{p}^\nu
  - \Gamma^i{}_{\alpha\beta}\hat{p}^\alpha\hat{p}^\beta \right]
  \frac{\partial}{\partial p^i}.
\end{equation}
This vector field is divergence-free on $(\Gamma_m,\hat{h})$, as the next theorem shows.

\begin{theorem}[Liouville's theorem, cf. \cite{jE73}]
\label{Thm:Liouville}
The Liouville vector field $L_F$, when restricted to $\Gamma_m$, satisfies
\begin{equation}
\divrg L_F = 0,
\end{equation}
where the divergence operator refers to the Lorentz manifold $(\Gamma_m,\hat{h})$.
\end{theorem}

\proof The proof is based on Lemma~\ref{Lem:VolumeTM} and proceeds along the same lines as in the uncharged case, see Theorem~1 in Ref.~\cite{oStZ13b}.
\qed

\subsection{Distribution function and associated current density}

As for the uncharged case, we introduce the distribution function $f:\Gamma_m^+ \toÊ\Real$, $f\geq 0$, and the associated current density
\begin{equation}
{\cal J}_F := f \frac{L_F}{m}\in {\cal X}(\Gamma_m^+).
\label{Eq:JCurrent}
\end{equation}
Like the Liouville vector field $L_F$, the current density acquires a vertical component in the charged case. Taking into account the expression in Eq.~(\ref{Eq:LF}), ${\cal J}_F$ can be decomposed as
\begin{equation}
{\cal J}_F = f\frac{L}{m} + f q (I^V)^{-1}\left(\tilde{F}(p/m) \right),
\label{Eq:JCurrentSplit}
\end{equation}
where $L = (I^H)^{-1}(p)$ is the Liouville vector field in the absence of charges, which is horizontal and satisfies $\hat{g}(L,L) = -m^2$. The splitting~(\ref{Eq:JCurrentSplit}) is analogous to what occurs in the familiar decomposition of the current density into advection and conduction currents, where the horizontal part $f L/m$ plays the role of the advection current. Note that the vertical part is proportional to the vertical lift of $\tilde{F}(p/m)$ where the latter is the electric field measured by the charged particles.

Given a smooth, $2d$-dimensional spacelike hypersurface $\Sigma$ in $\Gamma_m^+$ with unit normal vector field $\nu$, the flux integral
\begin{equation}
N[\Sigma] := -\int\limits_\Sigma \hat{h}({\cal J}_F,\nu) \eta_\Sigma
\label{Eq:NSigma}
\end{equation}
physically represents the ensemble average of occupied trajectories that intersect $\Sigma$, where here $\eta_\Sigma$ is the volume element on $\Sigma$ induced by the metric $\hat{h}$.\footnote{The minus sign comes from the timelike character of ${\cal J}_F$ and $\nu$, implying that $\hat{h}({\cal J}_F,\nu)$ is negative if they point in the same component of the light cone.} 

\begin{figure}[ht]
\centerline{\resizebox{4.0cm}{!}{\includegraphics{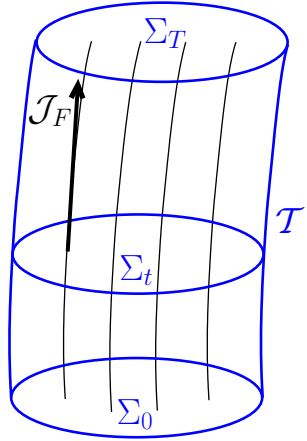}}}
\caption{The tubular region $V$ obtained by the flow of the spacelike compact hypersurface $\Sigma_0$ along ${\cal J}_F$. By construction ${\cal J}_F$ is tangent to ${\cal T}$ and thus its flow through ${\cal T}$ vanishes.}
\label{Fig:Tube}
\end{figure}

For a tubular region $V = \bigcup_{0\leq t \leq T}\Sigma_t$ which is obtained by letting flow along the integral curves of ${\cal J}_F$ a $2d$-dimensional, spacelike hypersurface $\Sigma_0$ in $\Gamma_m^+$, the boundary of $V$ consists of the initial and final hypersurfaces $\Sigma_0$ and $\Sigma_T$ and the cylindrical piece, ${\cal T} := \bigcup_{0\leq t\leq T}\partial\Sigma_t$, see Figure~\ref{Fig:Tube}. As a consequence of Gauss' theorem, we obtain
\begin{equation}
N[\Sigma_T] - N[\Sigma_0] = \int\limits_V (\divrg{\cal J}_F) \eta_{\Gamma_m}.
\label{Eq:BalanceLaw}
\end{equation}
The expression on the left-hand side of this equation is equal to the ensemble average of the net change in number of occupied trajectories between $\Sigma_0$ and $\Sigma_T$ due to collisions. For the particular case of a collisionless gas it follows from Eq.~(\ref{Eq:BalanceLaw}) that $\divrg{\cal J}_F = 0$. Taking into account the definition of ${\cal J}_F$ and Theorem~\ref{Thm:Liouville}, this implies that the distribution function $f$ must satisfy the Liouville equation
\begin{equation}
L_F[f] = 0.
\label{Eq:LiouvilleEq}
\end{equation}

\subsection{Observables derived from the distribution function}

The current density ${\cal J}_F$ on the mass shell $\Gamma_m^+$ gives rise to a natural, physically observable current $J$ on the spacetime. In order to explain this, we consider a $d$-dimensional spacelike hypersurface  $S$ in $(M,g)$ with unit normal vector field $s$. This surface can be lifted to a $2d$-dimensional hypersurface $\Sigma$ of $\Gamma_m^+$, defined as
\begin{equation}
\Sigma := \{ (x,p) : x\in S, p\in P_x^+ \} \subset \Gamma_m^+
\label{Eq:Sigma}
\end{equation}
with associated unit normal vector field $\nu := (I^H)^{-1}(s)$ (see Ref.~\cite{oStZ13b}). Now let us compute the ensemble average of occupied trajectories that intersect this hypersurface $\Sigma$. Assuming for simplicity that $S$ is entirely contained inside a coordinate neighborhood of $M$ we have, in adapted local coordinates,
\begin{displaymath}
\nu = (I^H)^{-1}(s) = s^\mu\hat{e}_\mu,
\end{displaymath}
and together with the expression~(\ref{Eq:VolumeFormMassShellLocalCoord}) this yields
\begin{equation}
\eta_\Sigma = i_\nu\eta_{\Gamma_m} = \frac{m}{\hat{p}_0}\det(g_{\mu\nu})
\frac{1}{d!} s^{\mu_0}\varepsilon_{\mu_0\mu_1\ldots\mu_d} 
dx^{\mu_1}\wedge\ldots\wedge dx^{\mu_d}\wedge dp^1\wedge\ldots\wedge dp^d
 = m(i_s\eta)\wedge \pi_x,
\label{Eq:LocalSplitting}
\end{equation}
where we have introduced the natural volume element $\eta$ on $(M,g)$, defined by
\begin{displaymath}
\eta := \sqrt{-\det(g_{\mu\nu})} dx^0\wedge dx^1\wedge\ldots\wedge dx^d,
\end{displaymath}
and the natural volume element on the future mass hyperboloid $P_x^+$, defined by
\begin{displaymath}
\pi_x := \frac{\sqrt{-\det(g_{\mu\nu})}}{-\hat{p}_0} dp^1\wedge\ldots\wedge dp^d.
\end{displaymath}
Using the local splitting~(\ref{Eq:LocalSplitting}), it follows that the integral of any sufficiently smooth and fast decaying function $F$ over a hypersurface $\Sigma$ of the form~(\ref{Eq:Sigma}) can be computed as
\begin{equation}
\int\limits_\Sigma F \eta_\Sigma 
 = m\int\limits_S \left( \int\limits_{P_x^+} F \pi_x \right) \eta_S,
\end{equation}
with $\eta_S := i_s\eta$ the induced volume element on $S$. This Fubini-type formula still holds true if the surface $S$ is not included in a coordinate chart, as can be shown by using a partition of unity. In particular, choosing $F(x,p) := -\hat{h}_{(x,p)}({\cal J}_F,\nu) = g_x(p,s)/m$ it follows that
\begin{equation}
N[\Sigma] = -\int\limits_\Sigma \hat{h}({\cal J}_F,\nu) \eta_\Sigma
 = -\int\limits_S g(J,s) \eta_S,
\label{Eq:JJ}
\end{equation}
where the current density $J\in {\cal X}(M)$ is defined as
\begin{equation}
J_x(\omega) := \int\limits_{P_x^+} f(x,p) \omega(p) \pi_x,\qquad
\omega\in T_x^*M.
\end{equation}
This vector field $J$ can be interpreted as the first moment of the distribution function in momentum space. Its flux integral through the hypersurface $S$ is equal to the flux integral of ${\cal J}_F$ through $\Sigma$ which is equal to $N[\Sigma]$. Obviously however, the vector field ${\cal J}_F$ on $\Gamma_m^+$ contains much more information than $J$ since it allows one to compute $N[\Sigma]$ for arbitrary $2d$-dimensional spatial hypersurfaces of $\Gamma_m^+$ and not just those which are of the form~(\ref{Eq:Sigma}). Note that in contrast to ${\cal J}_F$, the physically observable current $J$ does not depend on the electromagnetic field $F$. The reason for this relies in the fact that in the flux integral~(\ref{Eq:JJ}) $\nu$ is horizontal while the electromagnetic field only contributes to the vertical component of ${\cal J}_F$.

The next result shows that the divergence of $J$ is also related to the divergence of ${\cal J}_F$ through a fibre integral, and implies that $J$ is conserved.

\begin{proposition}
\label{Prop:ConservationLawJ}
Let $f: \Gamma_m^+\to\Real$ be a $C^\infty$-function of compact support on the future mass shell. Then, the following identity holds for all $x\in M$:
\begin{equation}
\divrg J_x = m\int\limits_{P_x^+} \divrg{\cal J}_F(x,p)\pi_x.
\label{Eq:DivJ}
\end{equation}
\end{proposition}

\proof As in the uncharged case, see Ref.~\cite{oStZ13b}.
\qed

As a consequence of Proposition~\ref{Prop:ConservationLawJ} and the fact that $m\divrg({\cal J}_F) = \divrg( f L_F) = L_F[f]$, we have
\begin{equation}
\divrg J_x = \int\limits_{P_x^+} L_F[f](x,p)\pi_x,
\label{Eq:DivJBis}
\end{equation}
implying that $J$ is conserved if the distribution function $f$ satisfies the Liouville equation $L_F[f] = 0$. 

The identity~(\ref{Eq:DivJBis}) can be generalized to arbitrary moments of the distribution function, defined by the following $s$-rank contravariant symmetric tensor field $T$ on $M$:
\begin{equation}
T_x(\omega^1,\ldots,\omega^s)
 := \int\limits_{P_x^+} f(x,p) \omega^1(p)\cdots\omega^s(p) \pi_x,
\qquad \omega^1,\ldots,\omega^s\in T_x^*M.
\label{Eq:TDef}
\end{equation}

\begin{proposition}
\label{Prop:ConservationLawT}
Let $f: \Gamma_m^+\to\Real$ be a $C^\infty$-function of compact support on the future mass shell. Then, the following identity holds for all $x\in M$ and all $\omega^2,\ldots,\omega^s\in T_x^*M$:
\begin{equation}
(\divrg T)_x(\omega^2,\ldots,\omega^s) 
 = \int\limits_{P_x^+} L_F[f](x,p)\omega^2(p)\cdots\omega^s(p)\pi_x 
 + q(s-1)\int\limits_{P_x^+} f(x,p)\omega^{(2}(\tilde{F}(p))\omega^3(p)\cdots
\omega^{s)}(p)\pi_x,
\label{Eq:DivT}
\end{equation}
where the round parenthesis denote total symmetrization, and where $\tilde{F}$ has been defined in Lemma~\ref{Lem:LF}.
\end{proposition}

\proof Fix $s-1$ one-forms $\omega^2,\ldots,\omega^s$ on $M$ and replace the function $f(x,p)$ with the function $A(x,p) := f(x,p)\omega^2_x(p)\cdots\omega^s_x(p)$ on both sides of Eq.~(\ref{Eq:DivJBis}), noticing that this procedure replaces $J$ with the vector field $T(\cdot,\omega^2,\ldots,\omega^s)$ on $M$. Then, using adapted local coordinates we obtain from Eq.~(\ref{Eq:DivJBis}),
\begin{displaymath}
\nabla_{\mu_1}\left( 
 T^{\mu_1\mu_2\ldots\mu_s}\omega^2_{\mu_2}\cdots\omega^s_{\mu_s}\right) 
 = \int\limits_{P_x^+} L_F[ f\omega^2(p)\cdots\omega^s(p) ]\pi_x,
\end{displaymath} 
or
\begin{eqnarray*}
&& \left( \nabla_{\mu_1} T^{\mu_1\mu_2\ldots\mu_s} \right)
\omega^2_{\mu_2}\cdots\omega^s_{\mu_s}
 + T^{\mu_1\mu_2\ldots\mu_s}\left( \nabla_{\mu_1}\omega^2_{\mu_2}\cdot
 \omega^3_{\mu_2}\cdots\omega^s_{\mu_s} + \ldots +
  \omega^2_{\mu_2}\cdots\omega^{s-1}_{\mu_{s-1}}
   \nabla_{\mu_1}\omega^s_{\mu_s} \right)\\
 &=& \int\limits_{P_x^+} L_F[f]\omega^2(p)\cdots\omega^s(p)\pi_x\\
 &+& \int\limits_{P_x^+} f(x,p) L_F[\omega^2(p)]\omega^3(p)\cdots\omega^s(p)\pi_x
 + \ldots
 + \int\limits_{P_x^+} f(x,p)\omega^2(p)\cdots\omega^{s-1}(p) L_F[\omega^s(p)] \pi_x.
\end{eqnarray*} 
Now the proposition follows by using the coordinate expression~(\ref{Eq:LLocCoord}) for $L_F$ and noticing that
\begin{displaymath}
L_F[\omega(p)] = p^\mu p^\nu\nabla_\mu\omega_\nu 
 + q F^\nu{}_\mu p^\mu\omega_\nu.
\end{displaymath}
\qed

As has been discussed in Refs.~\cite{jE71,oStZ13} the relevant observables for the Einstein-Maxwell-Vlasov system are the physically observable current density $J$ and the stress-energy tensor, corresponding to the tensor field $T$ defined in Eq.~(\ref{Eq:TDef}) with $s=2$. In adapted local coordinates these observables take the form:
\begin{equation}
J_x^\mu = \int\limits_{P_x^+} f(x,p) p^\mu \pi_x,\qquad
T_x^{\mu\nu} = \int\limits_{P_x^+} f(x,p) p^\mu p^\nu \pi_x.
\end{equation}
At any given event $x\in M$ for which $f (x,\cdot)$ is not identically zero, the current density $J_x$ is future-directed timelike. Furthermore, the stress-energy tensor satisfies the weak, the strong and the dominant energy conditions, see Lemma~7 in Ref.~\cite{oStZ13}.

%%%%%%%%%%%%%%%%%%%%%%%%%%%%%%%%%%%%%%%%%%%%%
\section{Symmetries of the distribution function}
\label{Sec:Symmetries}
%%%%%%%%%%%%%%%%%%%%%%%%%%%%%%%%%%%%%%%%%%%%%

In a previous article~\cite{oStZ13b}, see also Refs.~\cite{rMdT93,rMdT94}, we discussed in detail how to lift one-parameter groups of diffeomorphisms of the spacetime manifold $M$ to the tangent bundle $TM$. This lift induces a lift ${\cal X}(M)\to {\cal X}(TM), \xi\mapsto \hat{\xi}$ of the corresponding generator $\xi$. In particular, for symmetry groups $S$, that is, one-parameter groups of \emph{isometries} of ($M,g)$, it follows that the lifted group is an isometry group of $(TM,\hat{g})$. Using this lift, we define the distribution function $f$ to be $S$-symmetric if $f$ is invariant with respect to the lifted group, that is, if $\pounds_{\hat{\xi}} f = 0$. As we have discussed in~\cite{oStZ13b}, this definition is compatible with the imposition of the Liouville equation $\pounds_L f = 0$, since $[L,\hat{\xi}] = 0$ whenever $\xi$ is a Killing vector. Moreover, it also follows that the infinitesimal generator $\hat{\xi}$ is tangent to the mass shell $\Gamma_m$, and thus $\Gamma_m$ is invariant under the associated flow. Finally, it was shown that $\hat{\xi}$ generates symplectic transformations on $(TM,\Omega_s)$, a property that was exploited in the construction of explicit solutions of the Liouville equation on a Kerr black hole background.

In this section, we analyze the impact of the electromagnetic field upon the symmetry properties of the distribution function associated to a collisionless, charged gas. We show that the main properties described above remain satisfied provided the electromagnetic field $F$ is $S$-symmetric, that is, if $\pounds_\xi F = 0$.

For an arbitrary vector field $\xi\in {\cal X}(M)$ on the spacetime manifold $M$, the lifted field $\hat{\xi}\in {\cal X}(TM)$ can be defined as~\cite{oStZ13b}
\begin{equation}
\hat{\xi} := (I^H)^{-1}(\xi) + (I^V)^{-1}(\nabla_p\xi),
\label{Eq:xiDef}
\end{equation}
and in adapted local coordinates $(x^\mu,p^\mu)$ this expression reduces to
\begin{equation}
\hat{\xi} = \xi^\mu e_\mu + \nabla_p\xi^\mu \frac{\partial}{\partial p^\mu}
 =  \xi^\mu\frac{\partial}{\partial x^\mu}
 + p^\alpha\frac{\partial \xi^\mu}{\partial x^\alpha}\frac{\partial}{\partial p^\mu}.
\label{Eq:xihatLocCoord}
\end{equation}
The lift satisfies the following properties:

\begin{proposition}[\cite{sS58,oStZ13b}]
\label{Prop:Sym}
Let $\xi,\eta\in {\cal X}(M)$ be vector fields on $M$. Then, the lifted vector fields $\hat{\xi},\hat{\eta}\in {\cal X}(TM)$ satisfy
\begin{enumerate}
\item[(i)] $[\hat{\xi},\hat{\eta}] = \hat{\zeta}$ where $\zeta := [\xi,\eta]$.
\item[(ii)] $\hat{\xi}$ is a Killing vector field on $(TM,\hat{g})$ if and only if $\xi$ is a Killing vector field on $(M,g)$.
\item[(iii)] $dH_{(x,p)}(\hat{\xi}) = g_x(p,\nabla_p\xi)$ for all $(x,p)\in TM$.
\end{enumerate}
\end{proposition}

The first property shows that the lift preserves the commutator. Together with the second property this implies that Lie groups of isometries of $(M,g)$ lift to Lie-groups of isometries of $(TM,\hat{g})$. The third property means that if $\xi$ is Killing, then $dH(\hat{\xi}) = 0$ and thus $\hat{\xi}$ is tangent to $\Gamma_m$.

Let $\xi\in {\cal X}(M)$ be Killing vector field of the spacetime manifold $(M,g)$ which generates a one-parameter group $S$ of isometries, and let $\hat{\xi}\in {\cal X}(TM)$ be the corresponding lifted vector field on $(TM,\hat{g})$. We define the distribution function $f$ to be $S$-symmetric if it is invariant with respect to the flow generated by $\hat{\xi}$, that is, if
\begin{equation}
\pounds_{\hat{\xi}} f = 0.
\label{Eq:fSym}
\end{equation}
For the case that $f$ satisfies the Liouville equation for a charged gas, it follows from $L_F[f] = 0$ and Eq.~(\ref{Eq:fSym}) that
\begin{equation}
\pounds_{[L_F,\hat{\xi}]} f = 0,
\label{Eq:LSymConstraint}
\end{equation}
which is a constraint on $f$. A direct calculation using adapted local coordinates and the commutation relation~(\ref{Eq:ComRel}) reveals that
\begin{equation}
[L_F,\hat{\xi}] = p^\alpha p^\beta\left[Ê\nabla_\alpha\nabla_\beta\xi^\mu
 - R^\mu{}_{\alpha\beta\nu}\xi^\nu \right] \frac{\partial}{\partial p^\mu}
 - q(\pounds_\xi F^\mu{}_\alpha) p^\alpha\frac{\partial}{\partial p^\mu}.
\label{Eq:LSymCommutator}
\end{equation}
Whenever $\xi$ is a Killing vector field of $(M,g)$, the expression inside the square parenthesis vanishes. Therefore, for an uncharged gas, the Killing property of $\xi$ guarantees that the constraint~(\ref{Eq:LSymConstraint}) holds identically. However, for the charged case, the Killing property of $\xi$ is not sufficient for the vanishing of the commutator $[L_F,\hat{\xi}]$, it requires the additional condition of the vanishing of the Lie-derivative of $F$ with respect to $\xi$.

Finally, we show that the flow generated by $\hat{\xi}$ is symplectic if the corresponding flow on the spacetime manifold leaves both the metric and the electromagnetic field invariant.

\begin{proposition}
\label{Prop:Canonical}
Let $\xi\in {\cal X}(M)$ be the generator of a one-parameter group of isometries on $(M,g)$ which leaves the electromagnetic potential $A$ invariant. Then, the lifted vector field $\hat{\xi}$ satisfies the identity
\begin{equation}
i_{\hat{\xi}}\Omega_F = -dP,\qquad
P := \Theta_A(\hat{\xi}),
\end{equation}
where the one-form $\Theta_A$ is defined in Eq.~(\ref{Eq:PoincareOneForm}). In particular, $\hat{\xi}$ is the infinitesimal generator of a symplectic flow on $TM$, that is $\pounds_{\hat{\xi}}\Omega_F = 0$.
\end{proposition}

\proof Applying the interior derivative $i_{\hat{\xi}}$ on both sides of $\Omega_F = d\Theta_A$, we obtain
\begin{equation}
i_{\hat{\xi}}\Omega_F = \pounds_{\hat{\xi}}\Theta_A - d i_{\hat{\xi}}\Theta_A
 = \pounds_{\hat{\xi}}\Theta_A - dP,
\end{equation}
where we have used the Cartan identity in the first step. Using the coordinate expression
\begin{equation}
\Theta_A = (g_{\mu\nu} p^\nu + q A_\mu) dx^\mu
\end{equation}
and the explicit form for $\hat{\xi}$ given in Eq.~(\ref{Eq:xihatLocCoord}) a short calculation reveals that
\begin{displaymath}
\pounds_{\hat{\xi}}\Theta_A = (p^\nu\pounds_\xi g_{\mu\nu} + q \pounds_\xi A_\mu) 
dx^\mu,
\end{displaymath}
which vanishes according to our assumptions. Therefore, $i_{\hat{\xi}}\Omega_F = - dP$ follows. Applying the exterior derivative operator $d$ on both sides of this equation yields $\pounds_{\hat{\xi}}\Omega_s = 0$, which shows that $\hat{\xi}$ generates a symplectic flow on $TM$.
\qed

An important consequence of this proposition is that the quantity $P = \Theta_A(\hat{\xi})$ is conserved along the Liouville flow if $\xi$ is a Killing vector field of $(M,g)$ and at the same time $\pounds_\xi A = 0$. This follows from the well-known identity
\begin{equation}
dP(L_F) = \Omega_F(L_F,\hat{\xi}) =: \{ H,P \} = -\{ P,H \}
 = -\Omega_F(\hat{\xi},L_F) = -dH(\hat{\xi})
\end{equation}
which shows that $P$ is conserved along the Liouville flow if and only if the Poisson bracket $\{ H,P \}$ between $H$ and $P$ is zero which is the case if and only if $dH(\hat{\xi}) = 0$. The latter condition is a consequence of Proposition~\ref{Prop:Sym}(iii) and the fact that $\xi$ is Killing. Using standard arguments from Hamiltonian mechanics~\cite{Arnold-Book}, $\{ H,P \} = 0$ implies that the corresponding Hamiltonian vector fields $L_F$ and $\hat{\xi}$ commute with each other. This provides an alternative, more elegant explanation for the compatibility of the condition~(\ref{Eq:fSym}) with the Liouville equation $L_F[f] = 0$.

%%%%%%%%%%%%%%%%%%%%%%%%%%%%%%%%%%%%%%%%%%%%%
\section{Application: Collisionless distribution functions on a Kerr-Newman background}
\label{Sec:Applications}
%%%%%%%%%%%%%%%%%%%%%%%%%%%%%%%%%%%%%%%%%%%%%

As an application of our formalism, in this section, we discuss the Liouville equation $L_F[f] = 0$ on a Kerr-Newman background describing a charged, rotating black hole configuration. For the uncharged case, we showed in Ref.~\cite{oStZ13b} that the Liouville vector field can be trivialized by means of a suitable symplectic transformation on the tangent bundle. This symplectic transformation was constructed by finding a complete solution of the Hamilton-Jacobi equation which, as a consequence of the Killing structure of spacetime~\cite{mWrP70}, is separable~\cite{bC68}.

Here we generalize the method discussed in Ref.~\cite{oStZ13b} to the charged case. The main difference between the charged and the uncharged case lies in the fact that the vector $p$ in the definition of the Hamiltonian function $H(x,p)$ in Eq.~(\ref{Eq:HDef}) represents the \emph{physical} momentum which is different from the \emph{canonical} momentum $\pi$. In the charged case, the canonical momentum $\pi$ is given by the one-form
\begin{equation}
\pi := g(p,\cdot) + q A,
\end{equation}
where $A$ is the electromagnetic potential one-form. In terms of adapted local coordinates $(x^\mu,p^\mu)$ the symplectic form $\Omega_F$ defined in Eq.~(\ref{Eq:OmegaF}) is
\begin{equation}
\Omega_F = d\pi_\mu\wedge dx^\mu,\qquad \pi_\mu = g_{\mu\nu} p^\nu + q A_\mu,
\label{Eq:OmegaFKN}
\end{equation}
and $(x^\mu,\pi_\mu)$ are symplectic coordinates on $TM$. The Hamilton-Jacobi equation reads
\begin{equation}
g^{\mu\nu} p_\mu p_\nu = -m^2,\qquad 
p_\mu = \frac{\partial S}{\partial x^\mu} - q A_\mu,
\end{equation}
where $S: M\to \Real$ is the generating function on $(M,g)$. Parametrizing the Kerr-Newman metric in terms of Boyer-Lindquist coordinates (see, for instance, Refs.~\cite{MTW-Book,Heusler-Book}) this yields
\begin{eqnarray}
&-& \frac{1}{\Delta} \left[
 (r^2+a_H^2)\left( \frac{\partial S}{\partial t} - q A_t \right)
  + a_H\left( \frac{\partial S}{\partial\varphi} - q A_\varphi \right) \right]^2
 + \left[ \frac{1}{\sin\vartheta}\left( \frac{\partial S}{\partial\varphi} - q A_\varphi \right)
 + a_H\sin\vartheta\left( \frac{\partial S}{\partial t} - q A_t \right) \right]^2
\nonumber\\
&+& \Delta\left( \frac{\partial S}{\partial r} \right)^2 
 + \left( \frac{\partial S}{\partial\vartheta} \right)^2
 = -m^2\left(  r^2 + a_H^2\cos^2\vartheta \right),
\label{Eq:KerrNewmanHJ}
\end{eqnarray}
where the quantity $\Delta$ is defined as
\begin{displaymath}
\Delta := r^2 - 2m_H r + a_H^2 + q_H^2,
\end{displaymath}
the non-vanishing components of the electromagnetic potential are
\begin{displaymath}
A_t = \frac{q_H r}{r^2 + a_H^2\cos^2\vartheta},\qquad
A_\varphi = -\frac{q_H a_H r\sin^2\vartheta}{r^2 + a_H^2\cos^2\vartheta},
\end{displaymath}
and where $m_H$, $a_H$ and $q_H$ denote the mass, rotation parameter and charge of the black hole, respectively.

The two commutating Killing vector fields
\begin{displaymath}
k = \frac{\partial}{\partial t},\qquad
l = \frac{\partial}{\partial\varphi}
\end{displaymath}
generate isometries with respect to time translations and rotations, respectively, and their flows also leave the electromagnetic potential $A$ invariant. It follows from the remarks below Proposition~\ref{Prop:Canonical} that the quantities
\begin{eqnarray*}
E &:=& -\Theta_A(\hat{k}) = -p_t - q A_t = -\pi_t,\\
\ell_z &:=& \Theta_A(\hat{l}) = p_\varphi + q A_\varphi = \pi_\varphi
\end{eqnarray*}
are constant along the trajectories. Therefore, we make the following ansatz for the generating function:
\begin{equation}
S = -E t + \ell_z\varphi + S'(r,\vartheta).
\end{equation}
Substituting this into Eq.~(\ref{Eq:KerrNewmanHJ}) yields
\begin{equation}
\frac{1}{\Delta} \left[ (r^2+a_H^2)E - a_H\ell_z + q q_H r \right]^2 
 - m^2 r^2 - \Delta\left( \frac{\partial S'}{\partial r} \right)^2
= \left( \frac{\ell_z}{\sin\vartheta} - a_H\sin\vartheta E \right)^2 
 + m^2 a_H^2\cos^2\vartheta
 + \left( \frac{\partial S'}{\partial\vartheta} \right)^2,
\label{Eq:KerrNewmanHJSep}
\end{equation}
where we have used the fact that $a_H\sin\vartheta A_t + A_\varphi/\sin\vartheta = 0$ and $(r^2 + a_H^2)A_t + a_H A_\varphi = q_H r$. Remarkably, as noted in Ref.~\cite{bC68}, this equation is separable: If we set $S'(r,\vartheta) = S_r(r) + S_\vartheta(\vartheta)$ it follows that the left-hand side of Eq.~(\ref{Eq:KerrNewmanHJSep}) is a function of $r$ only, while the right-hand side is a function of $\vartheta$ only. Therefore, both sides are constant, and we obtain
\begin{eqnarray*}
\left( \frac{dS_r}{dr}\right)^2 &=& \frac{R(r)}{\Delta(r)^2},\quad
R(r) := \left[ (r^2+a_H^2)E - a_H\ell_z + q q_H r \right]^2 - \Delta(m^2 r^2 + \ell^2),\\
\left( \frac{dS_\vartheta}{d\vartheta} \right)^2 &=& \Theta(\vartheta),\quad
\Theta(\vartheta) := \ell^2 - \left( \frac{\ell_z}{\sin\vartheta} - a_H\sin\vartheta E \right)^2
  - m^2 a_H^2\cos^2\vartheta,
\end{eqnarray*}
where $\ell^2$ is the Carter constant.\footnote{Note that in the non-rotating limit, $\ell^2 = p_\vartheta^2 + p_\varphi^2/\sin^2\vartheta$, so $\ell$ is the total angular momentum in this case.} It follows that the generating function is, formally,
\begin{equation}
S(t,\varphi,r,\vartheta,m,E,\ell_z,\ell) = -E t + \ell_z\varphi
 + \int\limits^r \sqrt{R(r)} \frac{dr}{\Delta(r)}
 + \int\limits^\vartheta \sqrt{\Theta(\vartheta)} d\vartheta.
\end{equation}
According to the general theory of Hamilton-Jacobi~\cite{Arnold-Book}, this function generates a symplectic transformation
\begin{equation}
(t,\varphi,r,\vartheta,\pi_t,\pi_\varphi,\pi_r,\pi_\vartheta) \mapsto (Q^0,Q^1,Q^2,Q^3,P_0,P_1,P_2,P_3),
\label{Eq:SymplecticTransformation}
\end{equation}
where the new variables $(Q^a,P_a)$ are defined as
\begin{subequations}
\begin{eqnarray}
&& P_0 := m,\qquad P_1 := E,\qquad P_2 := \ell_z,\qquad P_3 := \ell,\\
&& Q^0 := \frac{\partial S}{\partial m} 
 = -m\int\limits^r\frac{r^2 dr}{\sqrt{R(r)}} 
 - m a_H^2\int\limits^\vartheta\frac{\cos^2\vartheta d\vartheta}{\sqrt{\Theta(\vartheta)}},\\
&& Q^1 := \frac{\partial S}{\partial E} 
 = -t + \int\limits^r\frac{ (r^2 + a_H^2)A(r)}{\sqrt{R(r)}}\frac{dr}{\Delta(r)}
 + a_H\int\limits^\vartheta\frac{B(\vartheta)}{\sqrt{\Theta(\vartheta)}}
 d\vartheta,\\
&& Q^2 := \frac{\partial S}{\partial \ell_z} 
= \varphi - a_H\int\limits^r\frac{A(r)}{\sqrt{R(r)}}\frac{dr}{\Delta(r)}
 - \int\limits^\vartheta\frac{B(\vartheta)}{\sqrt{\Theta(\vartheta)}}
 \frac{d\vartheta}{\sin^2\vartheta},\\
&& Q^3 := \frac{\partial S}{\partial \ell} 
 = -\ell\int\limits^r\frac{dr}{\sqrt{R(r)} }
 + \ell\int\limits^\vartheta\frac{d\vartheta}{\sqrt{\Theta(\vartheta)}},
\end{eqnarray}
\end{subequations}
with the functions $A(r) := (r^2 + a_H^2)E - a_H\ell_z + q q_H r$ and $B(\vartheta) := \ell_z - a_H\sin^2\vartheta E$. In fact, it can be easily verified that the transformation~(\ref{Eq:SymplecticTransformation}) leaves the symplectic form $\Omega_F$ in Eq.~(\ref{Eq:OmegaFKN}) invariant. Since $H = -m^2/2 = -P_0^2/2$, it follows from the Hamiltonian equations that the quantities $Q^1,Q^2,Q^3,P_0,P_1,P_2,P_3$ are constant along the trajectories and that
\begin{equation}
\dot{Q}^0 = \frac{\partial H}{\partial m} = -m.
\end{equation}
Consequently, the Liouville vector field in these new coordinates assumes the simple form
\begin{equation}
L_F = -m\frac{\partial}{\partial Q^0}.
\end{equation}
This offers an enormous simplification for the determination of a collisionless charged distribution function. In terms of the new coordinates the Liouville equation is equivalent to the statement that the distribution function has form:
\begin{equation}
f(x,p) = G(Q^1,Q^2,Q^3,P_0,P_1,P_2,P_3),
\label{Eq:fKerrNewman}
\end{equation}
for an arbitrary smooth function $G$. Finally, we notice that the natural lift defined in Eq.~(\ref{Eq:xiDef}) of the Killing vector fields $k$ and $l$ is simply
\begin{equation}
\hat{k} = -\frac{\partial}{\partial Q^1},\qquad
\hat{l} = \frac{\partial}{\partial Q^2}.
\end{equation}
Therefore, the distribution function $f$ in Eq.~(\ref{Eq:fKerrNewman}) is stationary if and only if $G$ is independent of $Q^1$, and it is axisymmetric if and only if $G$ is independent of $Q^2$.

It should be mentioned that our analysis has been formal. In general, the generating function $S$, and as a consequence also the functions $Q^0,\ldots,Q^3$ are multi-valued. Therefore, appropriate periodicity conditions on $G$ need to be specified in order for the distribution function to be well-defined. Specific applications will be discussed in future work.

%%%%%%%%%%%%%%%%%%%%%%%%%%%%%%%%%%%%%%%%%%%%%
\section{Conclusions}
\label{Sec:Conclusions}
%%%%%%%%%%%%%%%%%%%%%%%%%%%%%%%%%%%%%%%%%%%%%

In this work, using the geometrical structure that the tangent bundle acquires from the spacetime $(M,g)$, we developed the relativistic kinetic theory of a simple, collisionless, charged gas. It is perhaps remarkable to note that even though the theory is more complex due to the electromagnetic force exerted between the charged particle, nevertheless as far the geometrical description of the theory is concerned, it only required the modification of the symplectic form. Since moreover the Hamiltonian has the same form as for the uncharged case, the natural metric $\hat g$ on $TM$ leads to a free of ambiguities, natural integration theory on the associated mass shells. Accordingly, and in parallel to what occurs for the uncharged case, a relativistic simple charged gas from the perspective of the tangent bundle can be interpreted as the flow of an incompressible fictitious fluid in the mass shell with the distribution function playing the role of a particle density. Observables for the charged case are also defined as fiber integrals involving the distribution function. 
 
As for the uncharged case, the  bundle metric $\hat g$ offers the means to provide a useful connection between the symmetries of the background metric $g$, the symmetries of the electromagnetic field $F$ and the symmetries of the distribution function. Using this connection, we have constructed the most general collisionless distribution function on a Kerr-Newman black hole spacetime. This construction takes into account the spacetime symmetries of the Kerr-Newman metric and associated background elecromagnetic field combined with the separability of the Hamilton-Jacobi equation. These results extend in a natural way those obtained in Ref.~\cite{oStZ13b}.

%%%%%%%%%%%%%%%%%%%%%%%%%%%%%%%%%%%%%%%%%%%%%
%% BACKMATTER
%%%%%%%%%%%%%%%%%%%%%%%%%%%%%%%%%%%%%%%%%%%%%

\begin{theacknowledgments}
We thank Pierre Bayard for fruitful discussions regarding the geometry of the tangent bundle. This work was supported in part by CONACyT Grant No. 101353 and by a CIC Grant to Universidad Michoacana.
\end{theacknowledgments}

%%%%%%%%%%%%%%%%%%%%%%%%%%%%%%%%%%%%%%%%%%%%
% Create the reference section using BibTeX:
\bibliographystyle{aipproc}  
\bibliography{refs_kinetic}
%%%%%%%%%%%%%%%%%%%%%%%%%%%%%%%%%%%%%%%%%%%%

\end{document}